\documentclass{elsart3}

\usepackage{epsfig}

\def\plotty#1#2{\centering \leavevmode\epsfxsize=#1\columnwidth \epsfbox{#2}}

\usepackage{amssymb}
\usepackage{natbib}

\usepackage{graphics}

\newcommand{\gray}{$\gamma$-ray}

\newcommand{\adv}{Adv.\ Space Res.}
\newcommand{\al}{Astrophys.\ Lett.}

\newcommand{\ass}{Astrophys.\ Space\ Sci.}

\newcommand{\aap}{Astron.\ \& Astrophys.}
\newcommand{\apj}{Astrophys.\ J.}
\newcommand{\pr}{Phys.\ Rev.}
\newcommand{\jgr}{J.\ Geophys.\ Res.}
\newcommand{\prl}{Phys.\ Rev.\ Lett.}
\newcommand{\ssr}{Space\ Sci.\ Rev.}

\newcommand{\nim}{Nucl.\ Instr.\ Meth.}

\newcommand{\icrc}{ICRC}
\newcommand{\url}[1]{\tt}

\newcommand{\pubjournal}[6] {#1, #2, #3 #4, #5, #6}
\newcommand{\pubjournala}[6]{#1, #6 #2 #3, #4 #5}
\newcommand{\pubjournalb}[6]{#1, #6, #2 #3 #4, #5}
\newcommand{\pubproc}[6]{#1, #2, #3, #4, #5, #6}
\newcommand{\pubproca}[6]{#1, #2, #3, #4, #5, #6}

\begin{document}

\begin{frontmatter}



\title{Observations of the Li, Be, and B isotopes and
constraints on cosmic-ray propagation}


\author[1]{G. A. de Nolfo},
\author[2,1]{I. V. Moskalenko},
\author[3]{W. R. Binns},
\author[4]{E.R. Christian},
\author[5]{A. C. Cummings},
\author[5]{A. J. Davis},
\author[5,6]{J. S. George},
\author[3]{P. L. Hink},
\author[3]{M. H. Israel},
\author[5]{R. A. Leske},
\author[3]{M. Lijowski},
\author[5]{R. A. Mewaldt},
\author[5]{E. C. Stone},
\author[7]{A. W. Strong},
\author[1]{T. T. von Rosenvinge},
\author[8]{M. E. Wiedenbeck},
\author[5]{N. E. Yanasak}

\address[1]{NASA/Goddard Space Flight Center, Code 661, Greenbelt, MD 20771 USA}

\address[2]{HEPL/Stanford University, Stanford, CA 94305 USA}

\address[3]{Department of Physics and McDonnell Center for the Space
Sciences, Washington University, St. Louis, MO 63130 USA}

\address[4]{NASA Headquarters, Washington DC 20546, USA}

\address[5]{Space Radiation Laboratory, California Institute of
Technology, Pasadena, CA 91125 USA}

\address[6]{The Aerospace Corporation, M2/260, Los Angeles, CA
90009, USA}

\address[7]{Max-Planck-Institut f\"ur extraterrestrische Physik,
Postfach 1603, D-85740 Garching, Germany}

\address[8]{Jet Propulsion Laboratory, California Institute of
Technology, Pasadena, CA 91109 USA}


\begin{abstract}

The abundance of Li, Be, and B isotopes in galactic cosmic rays
(GCR) between E=50-200 MeV/nucleon has been observed by the Cosmic
Ray Isotope Spectrometer (CRIS) on NASA's ACE mission since 1997
with high statistical accuracy. Precise observations of Li, Be, B
can be used to constrain GCR propagation models. We find that a diffusive reacceleration
model with parameters that best match CRIS results (e.g. B/C,
Li/C, etc) are also consistent with other GCR observations. A
$\sim$15--20\% overproduction of Li and Be in the model
predictions is attributed to uncertainties in the production
cross-section data. The latter becomes a significant limitation to
the study of rare GCR species that are generated predominantly via
spallation.

\end{abstract}


\end{frontmatter}


\section{Introduction}

The enormous excess of the abundances of Li, Be, and B (LiBeB)
relative to C, N, and O in GCRs by a factor of $\sim$$10^{4}$
compared with solar system abundances is attributed to the
fragmentation of primary GCR nuclei, mainly C, N, and O, on
interstellar H and He atoms \citep{freier}. Observations of
secondary GCRs such as LiBeB thus provide strong constraints on
propagation models of GCRs within the Galaxy, since the production
of these light isotopes depends on the amount of matter traversed
during propagation.  B/C ratio is often used to tune propagation
parameters, such that the model predictions agree automatically
with B/C. Li and Be are particularly interesting since their
production depends not only on the interaction of CNO, but also on
``tertiary'' interactions in the ISM (e.g. B, Be $\rightarrow$
Li), and therefore may provide further restrictions on propagation
models. In particular, one might expect to observe a stronger
energy dependence of Li/C and Be/C ratios compared to the B/C
ratio. In addition, establishing the details of GCR propagation in
the Galaxy will be beneficial for other studies, such as searches
for signatures of exotic physics in GCR, spectra and origin of
galactic and extragalactic \gray\ background, nucleosynthesis, and
solar modulation \citep[e.g.,][]{M05}.

The Cosmic Ray Isotope Spectrometer (CRIS) on the NASA/ACE
spacecraft \citep{stone} measures the isotopic composition of
elements with 2$\le$$Z$$\le$30 in the energy range from
$\sim$30--500 MeV/nucleon since 1997 with unprecedented precision.
High statistics allows LiBeB to be studied over an extended energy
range ($\sim$50--200 MeV/nucleon) for the first time. We present
absolute intensities and relative elemental and isotopic
abundances of GCR LiBeB observed by CRIS during near solar minimum
conditions, and we compare these results with observations from
previous instruments.  We also discuss the implications of these
observations in the context of a GCR transport model,
GALPROP.  GALPROP has recently been
improved to predict LiBeB GCR isotopic abundances, including
re-examined fragmentation cross sections used as input to the
model. The precision of the data set
from CRIS is high enough \citep[e.g.,][]{denolfo2001} to require a
re-evaluation of the contributing uncertainties in the model
calculations, particularly the uncertainties in the isotopic
cross-sections.

\section{Intensities and Elemental Abundances}\label{flux}

The absolute intensities of Li, Be, B, and C as measured by CRIS
are shown in Fig.~\ref{intensity}a (solid circles).  The data set
considered for this study covers near solar minimum conditions
between 1998 Jan. 1 and 1999 Jan. 23. Excluding periods of intense
solar activity, the total time period corresponds to 311 days
\citep[for details see][]{denolfo2003}. The uncertainties shown in
Fig.~\ref{intensity}a for CRIS observations include both
statistical and systematic uncertainties. The statistical
uncertainties are typically small, ranging from 0.5\% for C to 5\%
for Li. The systematic uncertainties result from the various
correction factors within the calculation for the absolute
intensity such as uncertainties in the determination of the
geometry factor (2\%), spallation loss within the instrument
(1--5\%), and the tracking efficiency of the Scintillating Optical
Fiber Hodoscope (SOFT) \citep{stone} ($<$1\% for $Z$$>$5 and up to
6\% for Li).

\newcommand{\ptimes}{+\mkern-14mu \times}
\newcommand{\sidediamond}{\rotatebox[origin=c]{90}{$\blacklozenge$}}
\newcommand{\ellipse}{\begin{picture}(8,5)(-3,-3)\oval(8,5)[]\end{picture}}
\newcommand{\lowerdiamond}{\lozenge\mkern-11mu\lower1.5pt\hbox{\tiny
$\blacktriangledown$}}
\newcommand{\upperbox}{\square\mkern-12.5mu\lower-1.5pt\hbox{\begin{picture}(5,5)(0,0)\linethickness{2.5pt}\line(6,0){5}\end{picture}}}
\newcommand{\crossa}{{\begin{picture}(8,8)(-2,1)\linethickness{1pt}\put(-4,4){\line(6,0){8}}\line(0,6){8}\end{picture}}}
\newcommand{\crossb}{{\begin{picture}(6,10)(-3,2)\put(-2,5){\line(6,0){4}}\line(0,6){10}\end{picture}}}

\begin{figure*} \plotty{1}{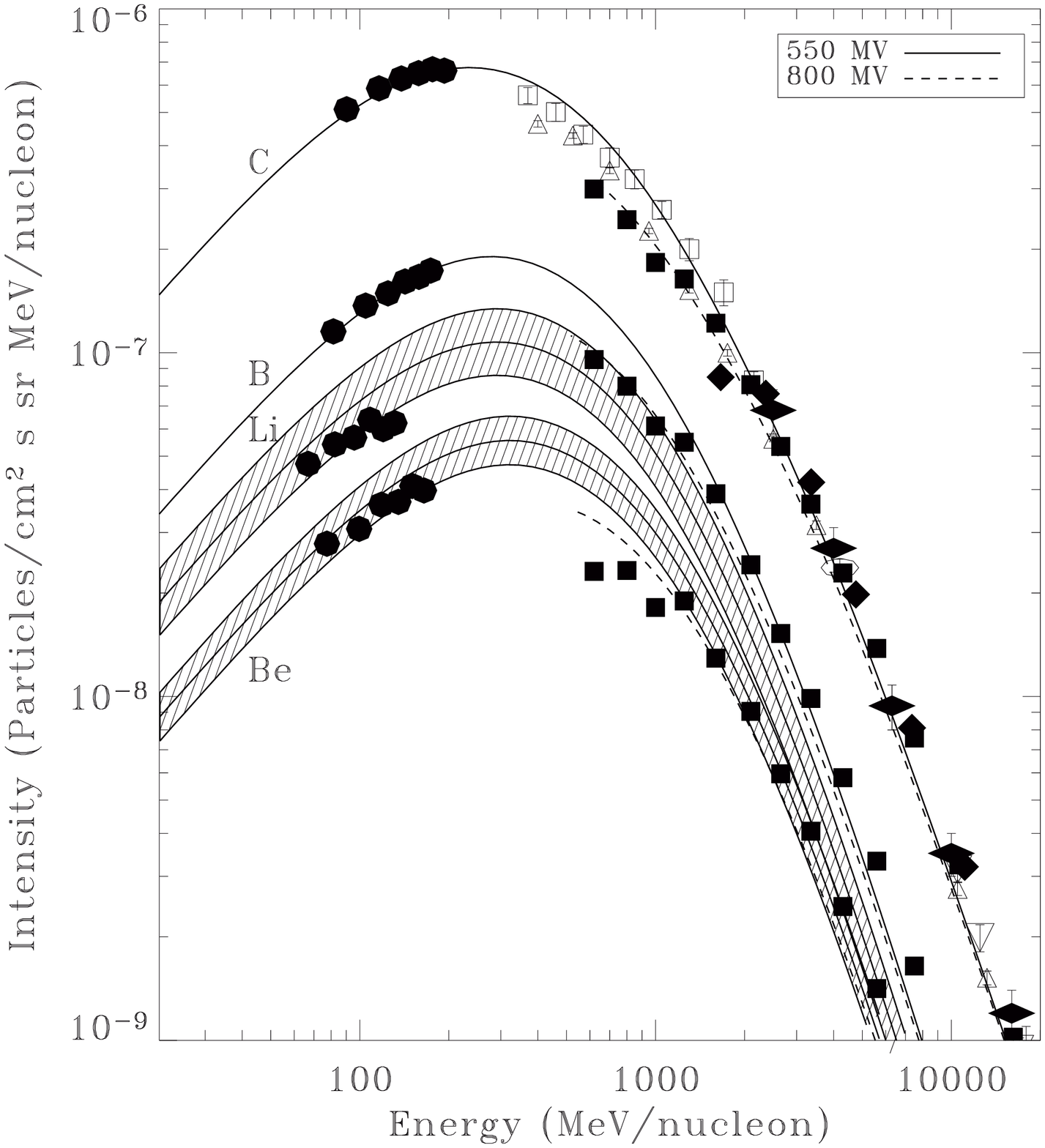}
\plotty{1}{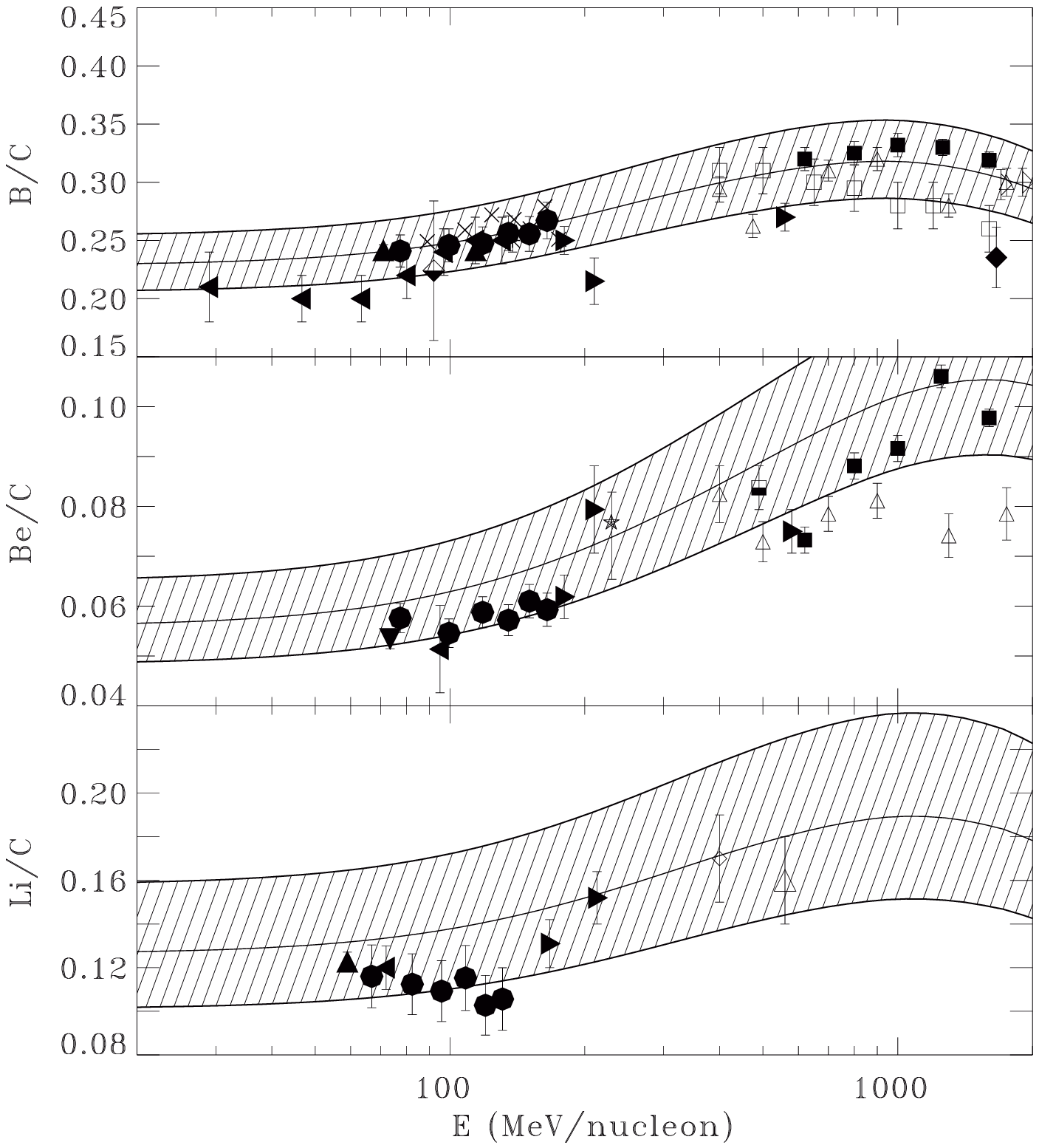} \caption[]{[a] Intensities of C,
B, Be, and Li shown for two levels of solar modulation ($\phi=550$
(solid curve) and 800 MV (dashed curve)) corresponding to the
epoch when the experimental data were taken, and [b] relative
elemental abundances compared with previous observations.  The
experimental data is compared with the results of a propagation
model GALPROP. See text for details. (Symbols refer to CRIS:
{\large $\bullet$}, Maehl et al.\ 1977: $\Box$, Webber et al.\
1972: {\normalsize $\triangle$}, Englemann et al.\ 1990:
$\blacksquare$, Orth et al.\ 1978: {\small $\blacklozenge$},
Buffington et al.\ 1978: {\small $\lozenge$}, Simon et al.\ 1980:
\sidediamond, Buckley et al.\ 1994: {\small \ellipse}, Chapel and
Webber 1981: {\small $\bigtriangledown$}, Lezniak et al.\ 1978:
{\small $\triangle$}, Muller et al.\ 1991: $\dagger$, Dwyer et
al.\ 1978: {\large $\triangleright$}, Webber et al.\ 1977:
$\blacktriangleright$, Garcia-Munoz et al.\ 1987:
$\blacktriangleleft$, Duvernois et al.\ 1996: $\ptimes$, Mewaldt
et al.\ 1981: $\lowerdiamond$, Wiedenbeck and Greiner 1980:
\crossa, Webber et al.\ 2002: $\blacktriangledown$, Hagen et al.\
1977: $\bigstar$, Fisher et al.\ 1976: {\small $\upperbox$},
Juliusson et al.\ 1974: {\bf +}, Garcia-Munoz et al.\ 1977:
{\large $\circ$}, Connell et al.\ 2001: {\normalsize $\times$},
Krombel et al.\ 1988: {\large $\triangleleft$}, Lukasiak et al.\
1999: $\blacktriangle$).} \label{intensity}
\end{figure*}

The elemental and isotopic ratios observed by CRIS are shown in
Fig.~\ref{intensity}b and Fig.~\ref{isotopes} for solar minimum
conditions. The ratios of the spallogenic nuclei (B, Be, and Li)
to mostly primary nuclei such as C are particularly important in
constraining propagation models since these ratios are sensitive
to the amount of material traversed by GCRs from the source to
detection at Earth. In addition, abundance ratios tend to be less
sensitive to instrumental uncertainties than absolute intensities.
In Fig.~\ref{intensity}b, the experimental error bars include both
systematic and statistical uncertainties.

In Fig.~\ref{intensity}, CRIS results are compared with previous
observations. A direct comparison of CRIS results with previous
observations is complicated at these energies by the effects of
solar modulation. The effect of solar modulation is approximated
by the spherically symmetric model of \citet{fisk}, characterized
by the solar modulation parameter $\phi$ \citep{gleeson}.  Levels
of modulation are determined for a given source spectrum by
matching post-propagation interstellar spectra as predicted by
GALPROP to CRIS at low energies and to HEAO--3 at high energies.
The
average amount of solar modulation experienced by particles
observed by CRIS during near-solar minimum conditions corresponds
to $\phi$$\approx$550 MV. While the modulation parameter can be
used to characterize the amount of modulation experienced during a
specific time period, the derived value of $\phi$ also depends on
the initial choice of the source spectrum and thus on the
propagated interstellar spectrum adopted as input in the solar
modulation calculation.  For instance, the solar modulation
adopted for CRIS during solar minimum is closer to $\phi$ =
$\sim$400 MV based on predictions from the Leaky Box Model
\citep{davis2001,niebur}.

CRIS observations are in good agreement with the previous
observations of Voyager 1, 2 and IMP 7, 8, both of which were made
during near solar modulation conditions similar to CRIS.  The
Voyager data were accumulated over 21 years with a weighted
average solar modulation level of $\phi$=450 MV.  This may not be
a good representation of the actual modulation level experienced
over 21 years \citep{connellbe10} and thus complicates a
comparison with CRIS data. At higher energies, the modulation
level determined for HEAO--3 data \citep{englemann1990}
corresponds to $\phi$$\sim$800 MV \citep{davis2000}.

\section{GCR Propagation Calculations}\label{model}

To interpret the CRIS observations of LiBeB spectra and isotopic
abundances, we use the diffusive reacceleration model \citep{seo}.
``Reacceleration'' is a distributed acceleration of particles due
to scattering on random hydromagnetic waves moving at Alfv\'en speed
in the interstellar
medium (Fermi 2nd-order mechanism). The model reproduces the 
peak near 1000 MeV/nucleon in the ratios of secondary to primary
nuclei in a physically motivated way (where the parameters are:
normalization and index of the diffusion coefficient, the Alfv\'en
speed, and the size of the galactic halo), and is consistent with
the K-capture parent/daughter nuclei ratio \citep{jones01a,
niebur}.

\begin{figure}[htb]
\includegraphics[
width=.42\textwidth]{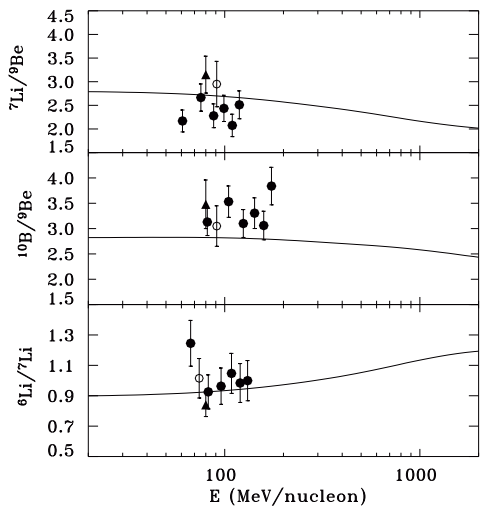}
\caption{Isotopic ratios of $^{6}$Li/$^{7}$Li, $^{10}$B/$^{9}$Be,
and $^{7}$Li/$^{9}$Be compared with previous data and with the
predictions of GALPROP. (Data from CRIS: {\large $\bullet$}, Garcia-Munoz
et al.\ 1977: {\large $\circ$}, Lukasiak et al.\ 1999:
$\blacktriangle$) }\label{isotopes}
\end{figure}

While the diffusive reacceleration model successfully reproduces
data on GCR nuclear species \citep{M02}, it underproduces GCR
antiprotons by a factor of $\sim$2 at 2 GeV, which may be a
signature of new effects. In particular, the propagation of
low-energy particles may be aligned to the magnetic field lines
instead of isotropic diffusion \citep{M02}, our local environment
(the Local Bubble) may produce a fresh ``unprocessed'' nuclei
component in GCR at low energy \citep{davis2000,M03}, or a more
intense nucleon spectra in distant regions could yield more
antiprotons and diffuse $\gamma$-rays \citep{SMR04a}. New accurate
data on LiBeB spectra and isotopic abundances by CRIS may be used
to test further the reacceleration model \citep{moskalenko03b}.

In our calculations we use the propagation model GALPROP as
described in detail elsewhere \citep{SM98,M02,SMR04a}. The model
is designed to perform GCR propagation calculations for nuclei
($Z$$\leq$28), antiprotons, electrons and positrons, and computes
$\gamma$-rays and synchrotron emission in the same framework.
GALPROP solves the transport equation with a given source
distribution and boundary conditions (free escape) for all GCR species.  This
includes a Galactic wind (convection), diffusive reacceleration,
energy losses, nuclear fragmentation, radioactive decay, and
production of secondary particles and isotopes.

The code includes cross-section measurements and energy dependent
fitting functions \citep{SM01}. The nuclear reaction network is
built using the Nuclear Data Sheets. The isotopic cross section
database is built using the extensive T16 Los Alamos compilation
of the cross sections \citep{t16lib} and modern nuclear codes
CEM2k and LAQGSM \citep{codes}. The most important isotopic
production cross sections ($^2$H, $^3$H, $^3$He, Li, Be, B, Al,
Cl, Sc, Ti, V, Mn) are calculated using our fits to major
production channels \citep[e.g.,][]{MMS01,M03,MM03}. Other cross
sections are calculated using phenomenological approximations by
\citet{W-code} (code {\tt WNEWTR.FOR} versions of 1993 and 2003)
and/or Silberberg and Tsao \citep{newyield} (code {\tt
YIELDX\_011000.FOR} version of 2000) renormalized to the data
where it exists. For $pA$ inelastic cross section we adapted the
parametrization by \citeauthor{crosec2} \citep[code
CROSEC,][]{crosec1,crosec2}.

The propagation equation is solved numerically starting at the
heaviest nucleus (i.e., $^{64}$Ni), computing all the resulting
secondary source functions, and then proceeds to the nuclei with
$A$--1.  The procedure is repeated down to $A$=1. To account for
some special $\beta^-$-decay cases (e.g., $^{10}$Be$\to$$^{10}$B)
the whole loop is repeated twice. The current version employs a full
3-dimensional spatial grid for all GCR species, but for the
purposes of this study the 2D cylindrically symmetrical option is
sufficient.

For a given size of Galactic halo, matching the propagation model
predictions to the observations of GCR B/C both at low energy with
CRIS data and at high energy with HEAO--3 data
\citep{englemann1990} determines the diffusion coefficient as a
function of momentum and the Alfv\'en speed. The halo size 4 kpc
is based on constraints set by the radioactive isotopes $^{10}$Be,
$^{26}$Al, $^{36}$Cl, and $^{54}$Mn \citep{SM01,MMS01}.  Assuming
a Kolmogorov spectrum of interstellar turbulence, it yields
$6.5\times10^{28}\beta(\rho/4\ {\rm GV})^{\delta}$ (cm$^{2}$
s$^{-1}$), where $\rho$ is the rigidity, $\delta$=1/3, the
Alfv\'en speed $v_A=35$ km s$^{-1}$, and the exact values of the
reacceleration parameters depend mainly on the adopted
cross-sections. The diffusion coefficient is assumed to be
independent on the spatial coordinates. While there is no a clear
evidence that the diffusion coefficient should be distinctly
different in the disk and in the halo, this minimizes the number
of parameters to be determined from the data.  The current state
of the data on radioactive isotopes do not warrant more a
complicated approach.  The measured isotopes to-date ($^{10}$Be,
$^{26}$Al, $^{36}$Cl, and $^{54}$Mn) all have halves lives around
1 Myrs.  To test the diffusion coefficient at different distances
from the Sun and in the halo, one needs accurate measurements of
radioactive isotopic abundances in CRs that include isotopes with
different lifetimes (e.g. $^{14}$C).  Heavy elements that have
large fragmentation cross sections may provide another way to test
the characteristics of the local interstellar medium \citep{M05}.

Supernova remnants (SNR) are believed to be the primary sources of
Galactic cosmic rays. Observations of X-ray \citep{Koyama1995} and
$\gamma$-ray emission \citep{Aharonian2005,Aharonian2006} from SNR
shocks reveal the presence of energetic particles thus testifying
to efficient acceleration processes. The predicted spectrum of
accelerated particles has power-law in rigidity with index which
may vary around $-2.0$ \citep[e.g.,][]{Ellison2005,Berezhko2006}.
Such a hard injection spectrum poses a difficulty in reconciling
the propagated spectrum with the direct CR measurements, assuming
the Kolmogorov spectrum of interstellar turbulence. The latter is
favored by the data \citep{Woo79,Sau99} and MHD simulations
\citep{Ver96}. In our calculations, the injection spectrum is
tuned to match the spectra of primary GCR nuclei for a given
propagation model. For the reacceleration model, the injection
spectrum is taken as a power-law in rigidity with index 2.1 below
9 GV and 2.42 above 9 GV, where the injection index at high
energies is fixed by the rigidity dependence of the diffusion
coefficient $\propto\rho^{1/3}$. The source abundances are tuned
to match the ACE/CRIS elemental and isotopic abundances
\citep{Wiedenbeck}.  Generally, the injection spectrum is model
dependent: different propagation models assume a different
rigidity dependence of the diffusion coefficient while the
propagated spectra are tuned to the same local data. For
reacceleration, this requires a change in the injection spectrum
index.  At low energies, direct measurements of the interstellar
spectrum are impossible due to the solar modulation, while
indirect observations of the proton spectrum via pionic gamma rays
by EGRET are not accurate enough and cannot be used to fix the
spectrum below a few GeV. The spectrum of CR nucleons below
$\sim$10GeV/nucleon is thus tuned to the local data assuming
modulation models, which are in turn, approximate and also rely on
the interstellar spectrum.

With the availability of precise GCR data from CRIS, uncertainties
in the fragmentation cross sections have become a significant
limitation to the study of rare GCR species that are generated
predominantly via spallation \citep{yanasak2001a,MMS01}. A review
of current cross sections \citep{MM03} was undertaken for the
dominant reactions involving the production of LiBeB as well as
products decaying to LiBeB species (e.g., $^6$He, $^{10,11}$C).
For instance, $\beta^-$-decay of $^{10}$Be contributes
significantly to $^{10}$B, while $^{7}$Be may produce some
$^{7}$Li via electron capture. \citet{sist1997} have surveyed
partial $^{7,10}$Be production cross sections for fragmentation of
O on hydrogen over a range of energies $E$$\sim$30--500
MeV/nucleon. Higher energy cross-section measurements
\citep[$E$$\sim$365--600 MeV/nucleon, from][]{W-code,webb1998} for
p+CNO$\rightarrow$LiBeB reactions have also been made.
\citet{michel1995} gives $^{7,10}$Be production cross sections for
fragmentation of CNO on hydrogen over a range of energies
$E$$\sim$800--2600 MeV/nucleon. In addition, measurements for
He+CN$\rightarrow$BeB reactions at 600 MeV/nucleon \citep{W-code}
and $\alpha+\alpha\to$ Li, $^7$Be at 60--160 MeV \citep{merc1997}
are available.

Li, Be, and B in cosmic rays are mainly produced from the
interaction of CNO nuclei with interstellar hydrogen. Some of the
dominant reactions, such as $^{12}$C $\to$ $^{6}$Li,
$^{7,9,10}$Be, $^{11}$B, $^{14}$N $\to$ $^{7}$Be, $^{16}$O $\to$
$^{6}$Li, $^{7,10}$Be, $^{9,11}$B, are constrained by the data.
The cross sections for the production of boron are better known
than for the lighter isotopes of beryllium and lithium, since the
main contribution to boron production is from CNO and $^{11}$B
$\to$ B.  Thus the model parameters are rather tightly constrained
by the measured B/C ratio at low and high energies.  However, some
channels of Be and Li show a spread in the data between 10-20\% or
greater. For example, $^{7}$Be production by CNO nuclei shows a
spread by a factor of $\sim$2 at energies below a few 100
MeV/nucleon. Some important channels are represented by one or few
data points in a narrow energy range (e.g., $^{11}$B, C, O $\to$
$^{10}$B, $^{11}$B $\to$ $^{9}$Be), while some are not measured at
all (e.g., $^{10}$B, N $\to$ $^{9}$Be, B, Be $\to$ $^{7}$Be). The
production of Li isotopes is particularly poorly measured. There
are only a few data points for Li production by CNO nuclei. The
data on Li isotopic production by $^{7}$Li, Be, B nuclei are
absent, while a contribution from spallation of these isotopes may
exceed $\sim$35\%. See \citet{MM03} for more details.

When there is no data, the production cross sections are
calculated using the semi-empirical parametrizations. This
translates to about a 15\% overall uncertainty in the isotopic
production ($\sim$20\% for Li), given that typical cross section
uncertainties of semi-empirical parametrizations are of the order
of $\sim$50\%.


The solid curves in Figs.~\ref{intensity} and \ref{isotopes}
correspond to GALPROP predictions modulated assuming a modulation
parameter of 550 MV, corresponding to the time period covered in
this study (Jan. 1, 1998 to Jan. 23, 1999). The dashed curves in
Fig.~\ref{intensity}a refer to modulated GALPROP predictions with
$\phi$=800 MV, appropriate for the modulation level during HEAO--3
observations.  The model is tuned to match the observations of C
and B intensities and the B/C ratio from CRIS
(Fig.~\ref{intensity}a/b). However, Be and Li intensities and the
ratios Li/C and Be/C are $\sim$10--20\% lower than predicted by
the model, although still consistent within the uncertainty
determined for the propagation calculation (shown as the hatched
region). The isotopic ratios shown in Fig.~\ref{isotopes} are also
in good agreement with the model predictions.  The model
uncertainties are mostly due to uncertainties in the cross
sections of tertiary interactions such as the production of Be and
Li from isotopes of Li, Be, and B as discussed above.  Further
work is required to pin down these uncertainties.

\section{Conclusion}

We have made GCR LiBeB abundance measurements using the CRIS
instrument during near solar minimum conditions in 1998--1999. The
isotopic ratios of LiBeB are in agreement with previous
measurements of GCR light isotopes, particularly for those
experiments acquiring data during periods of solar modulation
levels similar to the modulation levels in 1998--1999 covered by
CRIS.  The reacceleration model gives a satisfactory prediction
for GCR primary and secondary species with $Z$$\ge$3, and shows
good agreement with relative isotopic abundances
\citep{jones01b,M02}. In some cases, notably
($p,\alpha$)+(Li,Be,B)$\rightarrow$Be,Li, a lack of cross-section
measurements limits the reliability of model predictions. Understanding the subtle
differences between the model predictions and experimental data
now hinges on more precise cross section measurements, especially
for reactions involving the production of the light isotopes of
LiBeB.

This research was supported by NASA at the NASA/Goddard Space
Flight Center, the California Institute of Technology, (grant
NAG5-6912), the Jet Propulsion Laboratory, and Washington
University. I.V.M. acknowledges partial support from
NASA Astrophysics Theory Program (ATP) grant and NASA
Astronomy and Physics Research and Analysis Program (APRA) grant.


\end{document}